\documentclass[reprint,aip,jcp]{revtex4-1}

\usepackage{placeins}
\usepackage{dcolumn}
\usepackage{bm}
\usepackage[]{placeins}
\usepackage{graphicx}
\usepackage{caption}
\usepackage{subcaption}
\usepackage{amsmath}
\usepackage{natbib}

\newcommand{\kB}{k_\mathrm{B}}

\newcommand{\av}[1]{\langle #1 \rangle}
\newcommand{\dx}[1]{\text{d} #1}

\newcommand{\eff}{\mathrm{eff}}
\newcommand{\app}{\mathrm{app}}

\newcommand{\eq}[1]{Eq.~(\ref{#1})}
\newcommand{\eqs}[1]{Eqs.~(\ref{#1})}
\newcommand{\fig}[2]{Fig.~\ref{#1}#2}
\newcommand{\epsc}{\epsilon_\mathrm{c}}

\newcommand{\Tc}{T_\text{c}}

\begin{document}
 \title{Understanding the Nonlinear Dynamics of Driven Particles in Supercooled Liquids in Terms of an Effective Temperature}
\date{\today}
\author{Carsten F. E. Schroer}
\email{c.schroer@uni-muenster.de}
\author{Andreas Heuer}
\email{andheuer@uni-muenster.de}
\affiliation{Westf\"alische Wilhelms-Universit\"at M\"unster, Institut f\"ur physikalische Chemie, Corrensstra\ss e 28/30, 48149 M\"unster, Germany}
\affiliation{NRW Graduate School of Chemistry, Wilhelm-Klemm-Stra\ss e 10, 48149 M\"unster, Germany}

\begin{abstract}
{In active microrheology the mechanical properties of a material are tested by adding probe particles which are pulled by
an external force. In case of supercooled liquids, strong forcing leads to a thinning of the host material which becomes more pronounced as the system approaches the glass transition. In this work we provide a quantitative theoretical description of this thinning behavior based on the properties of the Potential Energy Landscape (PEL) of a model glass-former. A key role plays the trap-like nature of the PEL. We find that the mechanical properties in the strongly driven system behave the same as in a quiescent system at an enhanced temperature, giving rise to a well-characterized effective temperature. Furthermore, this effective temperature turns out to be independent of the chosen observable and individually shows up in the thermodynamic and dynamic properties of the system. Based on this underlying theoretical understanding, we can estimate its dependence on temperature and force by the PEL-properties of the quiescent system. We furthermore critically discuss the relevance of effective temperatures obtained by scaling relations for the description of out-of-equilibrium situations.}
\end{abstract}

\maketitle

\begin{figure*}
\begin{minipage}{0.45\textwidth}
\begin{flushleft}
 \textbf{(a)}
\end{flushleft}
\includegraphics[width=0.6\textwidth]{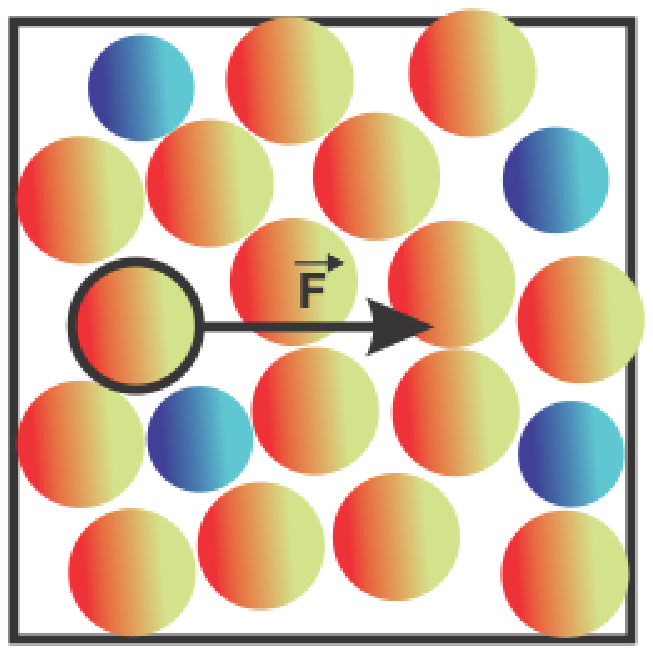}
\end{minipage}
\begin{minipage}{0.45\textwidth}
\begin{flushleft}
 \textbf{(b)}
\end{flushleft}
 \includegraphics[width=\textwidth]{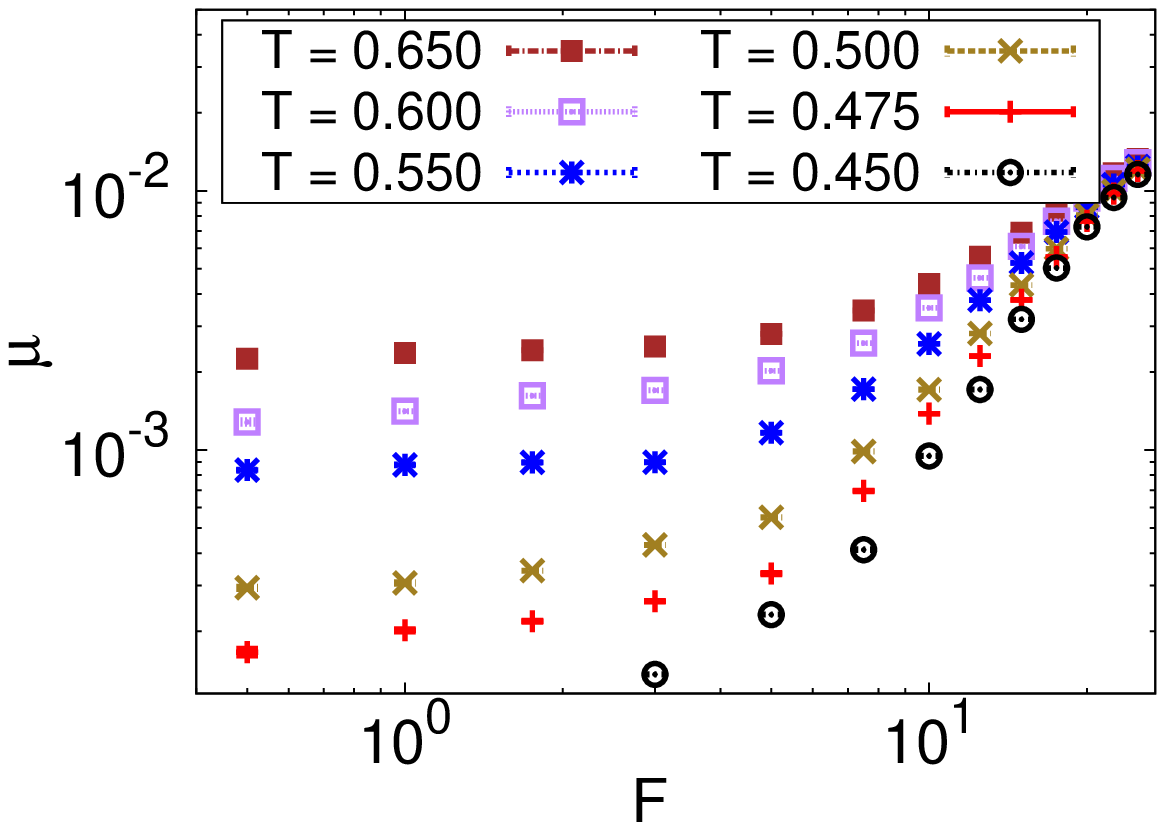}
\end{minipage} 

\caption{Active microrheology. (a) Schematic picture of active microrheology of a binary molecular liquid. (b) Mobility $\mu$ of the tracer particle as a function of the force $F$ for different temperatures $T$.}
\label{fig:rheology}
\end{figure*}

\section{Introduction}

In microrheology, the mechanical properties of a substance are studied by monitoring the trajectories of single particles \cite{Waigh2005,Cicuta2007, Squires2010, Puertas2014}. Thereby, one distinguishes between two different modes: Passive microrheology, in which the equilibrium properties of the material are probed by studying the motion of a single particle \cite{Mason1995,Waigh2005}, and active microrheology, in which a single particle is subjected to an external drive. In the latter case, the response to strong external perturbations is of particular interest since it allows one to study the properties of the system in an out-of-equilibrium situation \cite{Puertas2014}.

Recently, this method has been applied to study the properties of glass-forming systems in both, experiments \cite{Habdas2004,Wilson2009,Wilson2011} and simulations \cite{Reichhardt2006, Gazuz2009,Wilson2011,Winter2012,Ladadwa2013,Schroer2013,Schroer2013a,Reichhardt2015}.
For these systems, several remarkable effects have been observed like a nonlinear thinning of supercooled liquids \cite{Gazuz2009,Wilson2009,Wilson2011,Winter2012,Schroer2013}, local melting of glassy samples~\cite{Habdas2004,Voigtmann2013} and intermittent superdiffusivity in the supercooled regime~\cite{Winter2012,Schroer2013a}. Because of this, there is recently a field of vivid research to provide a theoretical description of the out-of-equilibrium properties \cite{Puertas2014}, mostly in the framework of mode-coupling theory \cite{Gazuz2009,Gazuz2013,Voigtmann2013}. For diluted colloidal solutions, a theoretical description in terms of the two-particle correlations between the probe and the host liquid has been developed by Brady et al.\cite{Squires2005,Carpen2005,Khair2006,Zia2010}.

A standard experiment in the context of active microrheology is the pulling of a single probe particle through a supercooled liquid with a constant external force $F$ \cite{Habdas2004,Winter2012,Winter2013,Schroer2013,Schroer2013a}, see \fig{fig:rheology}{
(a)} of a schematic visualization. Thereby, a characteristic quantity is the mobility $\mu = v/F$ of the probe, which is defined as the ratio of the stationary velocity $v$ of the particle along the force direction and the strength of the external field $F$. The mobility is determined by the resistance the particle experiences during the motion through the supercooled liquid and is therefore a direct measure for the mechanical properties of the system \cite{Squires2010}. As it can be seen in \fig{fig:rheology}{(b)}, the mobility is constant for small forces, which is in accordance with the Einstein relation \cite{Squires2010}

\begin{equation}
 \mu(F\to 0) = \frac{D}{\kB T},
\end{equation}
where $D$ denotes the equilibrium diffusion constant, and $\kB T$ the thermal energy of the system. For large forces, the mobility becomes itself force-dependent, thereby indicating a thinning of the local environment.

For the physical description of non-equilibrium states, one often finds the concept of \emph{effective temperatures} in literature \cite{Cugliandolo1997,Kob2000,Sciortino2001,Berthier2002,Ono2002,Lacks2004,Ilg2007,Wilson2011,Winter2012,Winter2013}. Thereby, a new temperature $T_\eff$ is defined which includes, on the one hand, the temperature imposed by the thermostat and, on the other hand, the control parameter of non-equilibrium, e.g. the maximum strain of a shear cycle or the waiting time in an aging system. The exact definition can result from evaluating the fluctuation-dissipation theorem \cite{Cugliandolo1997,Berthier2002,Ono2002}, studying the population of energies \cite{Kob2000,Sciortino2001,Lacks2004}, applying weakly interacting degrees of freedom that act as a thermometer \cite{Ilg2007,Wilson2011} or simply via scaling relations of different dynamic quantities \cite{Winter2012,Winter2013}. For the case of force-driven active microrheology the latter relations have been observed for the inverse mobility $\zeta$ and for the perpendicular diffusion coefficient of the tracer particle $D_\perp$, however, with quite different values of $T_\eff$. Since one would expect a ``physical'' temperature to be independent from the studied quantity, the insights these scaling temperatures are somewhat limited.

A novel way to describe microrheology in supercooled liquids is the application of the Potential Energy Landscape (PEL) approach. The basic idea of this approach is to relate the dynamics of the system to its thermodynamic state, characterized by its overall potential energy \cite{Goldstein1969,Stillinger1995}. In particular it turns out that the dynamics of supercooled liquids can be related to escape events out of mesoscopic funnels in the PEL, formed by a set of local minima, which is called Metabasins (MB) \cite{Stillinger1995}. Although the escape out of a single MB is a complicated process, which may contain several transitions between local minima, it can be described as a single activated process over a energy barrier that depends on the depth of the funnel \cite{Doliwa2003b}. This is in striking accordance to the trap model \cite{Monthus1996} that has been successfully applied to understand the connection between thermodynamics and dynamics in supercooled liquids \cite{Heuer2005}. On a more coarse grained level, the sequence of escapes from the MBs is largely uncorrelated and can be described in terms of a continuous time random walk (CTRW) scheme \cite{Doliwa2003c,Rubner2008}. 

Recently, the PEL approach has been used to describe the dynamics of a driven tracer particle in terms of a CTRW through its underlying PEL \cite{Schroer2013,Schroer2013a}. Thereby it has been found that the PEL approach offers a non-trivial decomposition of the linear and nonlinear response of $\mu$ into two independent quantities. In particular, one finds that the spatial part of the mobility behaves essentially constant, whereas the average waiting time $\av{\tau(F,T)}$ in between two transitions or, equivalently, the average rate $\av{\Gamma(F,T)}_P$ of a transition strongly varies; thereby, $\av{\dots}$ denotes a number average, whereas $\av{\dots}_P$ denotes the average taken with respect to the Boltzmann distribution of energies, see \eq{eq:defAvGamma} further below. This very different behavior means that the temporal part of the mobility is the main source of nonlinearity and, thus, that the understanding of the nonlinearity of the relevant timescale is identical to understanding the nonlinearity of the mobility itself.

In this article we offer a quantitative microscopic description of the thinning behavior of force-dependent mobility $\mu$. For this purpose we have conducted computer simulations of a typical glass-forming model liquid which allows us to study the real space dynamics simultaneous to the dynamics of the system in its underlying PEL (for details, see Methods). We first analyze the effect of the external force on the energy distribution, which can be rationalized in terms of an effective temperature $T_\eff$. For this effective temperature, a universal scaling with a single parameter is observed which can be applied for nearly all forces and temperatures studied in this work. Second, we study the force dependence of the local escape rates and provide a description in terms of an extended trap model. Thereby we find that the force dependence of this rates is governed by the same effective temperature as it was found for the thermodynamic quantity. Third, both quantities can be combined to obtain a quantitative description of the average escape rate $\av{\Gamma(F,T)}_P$ in terms of $T_\eff$. As a consequence the dynamics of the system behaves during the application of an external force similar to the dynamics of a quiescent system at a higher (bath) temperature, resulting in a Force-Temperature Superposition. This result can be directly applied to the real-space mobility. In the last step, we apply the concept, within some approximations though, to the diffusive properties of the system. Thereby we observe a similar scaling with the same effective temperature as for the mobility, however, the quality of the scaling depends on the studied quantity, due to the different validity of the applied approximation. We critically discuss that a slightly different choice of the scaling factor of the effective temperature can enhance the scaling properties of the diffusion coefficient. However, this can also lead to a loss of the physical meaning of the so defined effective temperature.

\section{Methods}

\begin{figure*}
 \includegraphics[width=0.45\textwidth]{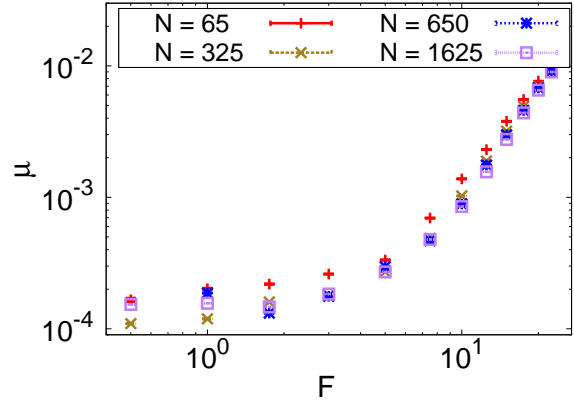}
\caption{Finite size effects. Mobility $\mu$ of the tracer particle as a function of the external force $F$ for systems with a different number of particles $N$ at a temperature $T=0.475$.}
\label{fig:fsize}
\end{figure*}

We have conducted molecular dynamics simulations (MD) of a binary (80:20) mixture Lennard-Jones particles where the potential parameters of Kob and Andersen~\cite{Kob1995} have been used. Since the analysis of the underlying PEL requires small system sizes~\cite{Buechner1999} the potential cutoff radius has been reduced to $1.8$ (in dimensionless Lennard-Jones units) which allows us to perform simulations of systems with $N=65$ particles. This system size is known to display only very small finite size effects concerning its diffusive behavior in equilibrium~\cite{Doliwa2003a,Rehwald2010a}. As is can be seen in \fig{fig:fsize}{}, also the mobility studied in this work is only very weakly dependent on the size of the system and the small system only displays deviations about $10-20\%$.

At the beginning of the simulation one of the majority particles of the system is randomly chosen and pulled by a constant force along a certain direction, thereby reaching a steady-state regime~\cite{Schroer2013}. Constant temperature conditions have been achieved by coupling the bath particles (all particles except of the tracer) to a Nos\'e-Hoover thermostat~\cite{Nose1984a,Hoover1985,Martyna1992}. We show in the Supplemental Material \cite{supp} that, in the range of forces we are studying in this work, the driving of the tracer causes no unphysical fluctuations of the kinetic energy of the bath particles. We study a (bath) temperature range of $T=0.45-0.65$ which is slightly above the mode-coupling temperature $T_\text{c}=0.435$ of the system~\cite{Kob1995}. In this regime neither incomplete structural relaxation nor aging has been observed.

The key step of our analysis is the tracking of the path the system has taken through its PEL. For this purpose we compute the current inherent structure of the system at certain points in time by minimizing the potential energy of the system via conjugate gradient method. The minimization is done with a copy of the system so that the MD trajectories are not influenced. For the computation of the potential energy, only the Lennard-Jones interactions are taken into account. This choice is necessary to compare the states of the driven and to those of the undriven system, since only in this case they all belong to the very same PEL.

At the end of the simulations, all inherent structures, between which forward-backwards transitions are observed, are combined to metabasins, whereas the structure with the lowest potential energy, which is typically the most frequently visited one, defines the energy. The exact protocol to determine the metabasins involves some minor subtleties which are described in~\cite{Doliwa2003b}.

\section{Results}

In the PEL description, the overall escape rate from an arbitrary MB depends on two formally independent quantities: The distribution $P(\epsilon,T)$ of MB energies $\epsilon$ populated by the system, which is a purely thermodynamic quantity and the rates $\av{\Gamma_\epsilon(\epsilon,T)}$ to escape from a MB of energy $\epsilon$, which describe the local kinetics.
The escape rates $\av{\Gamma_\epsilon(\epsilon,T)}$ are defined as the inverse average waiting time $1/\av{\tau_\epsilon(\epsilon,T)}$ before leaving a MB of energy $\epsilon$. The brackets thereby denote the average over different realizations. The total average escape rate or, equivalently, the inverse average waiting time is given by~\cite{Doliwa2003c,Heuer2008}

\begin{equation}
 \frac{1}{\av{\tau(T)}} = \av{\Gamma(T)}_P = \int{\dx{\epsilon}~\av{\Gamma_\epsilon(\epsilon,T)} P(\epsilon,T)}.
  \label{eq:defAvGamma}
\end{equation}

where the index $P$ of $\av{\Gamma(T)}_P$ shall emphasize that the average has been taken with respect to the energy distribution $P(\epsilon,T)$. Applying an external perturbation may alter the thermodynamics, the local kinetics or both in a different manner, giving rise to the observed nonlinear behavior of $\av{\Gamma(F,T)}_P$. Thus, the understanding of the force dependences of $\av{\Gamma_\epsilon(\epsilon,F,T)}$ and $P(\epsilon,F,T)$ immediately offer the understanding of the nonlinear mobility in general.

\subsection{Distribution of energies}

\begin{figure*}
\begin{minipage}{0.45\textwidth}
\begin{flushleft}
 \textbf{(a)}
\end{flushleft}
\includegraphics[width=\textwidth]{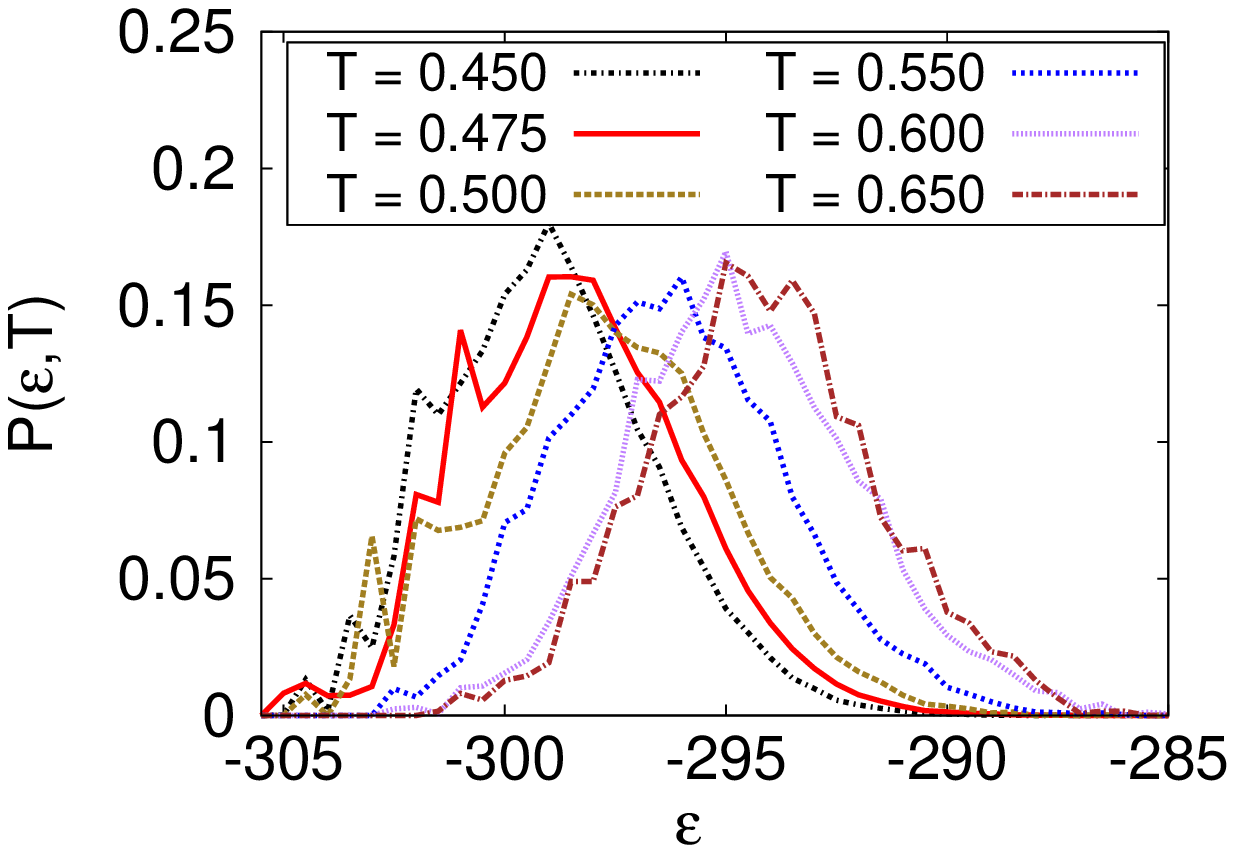}
\end{minipage}
\begin{minipage}{0.45\textwidth}
\begin{flushleft}
 \textbf{(b)}
\end{flushleft}
 \includegraphics[width=\textwidth]{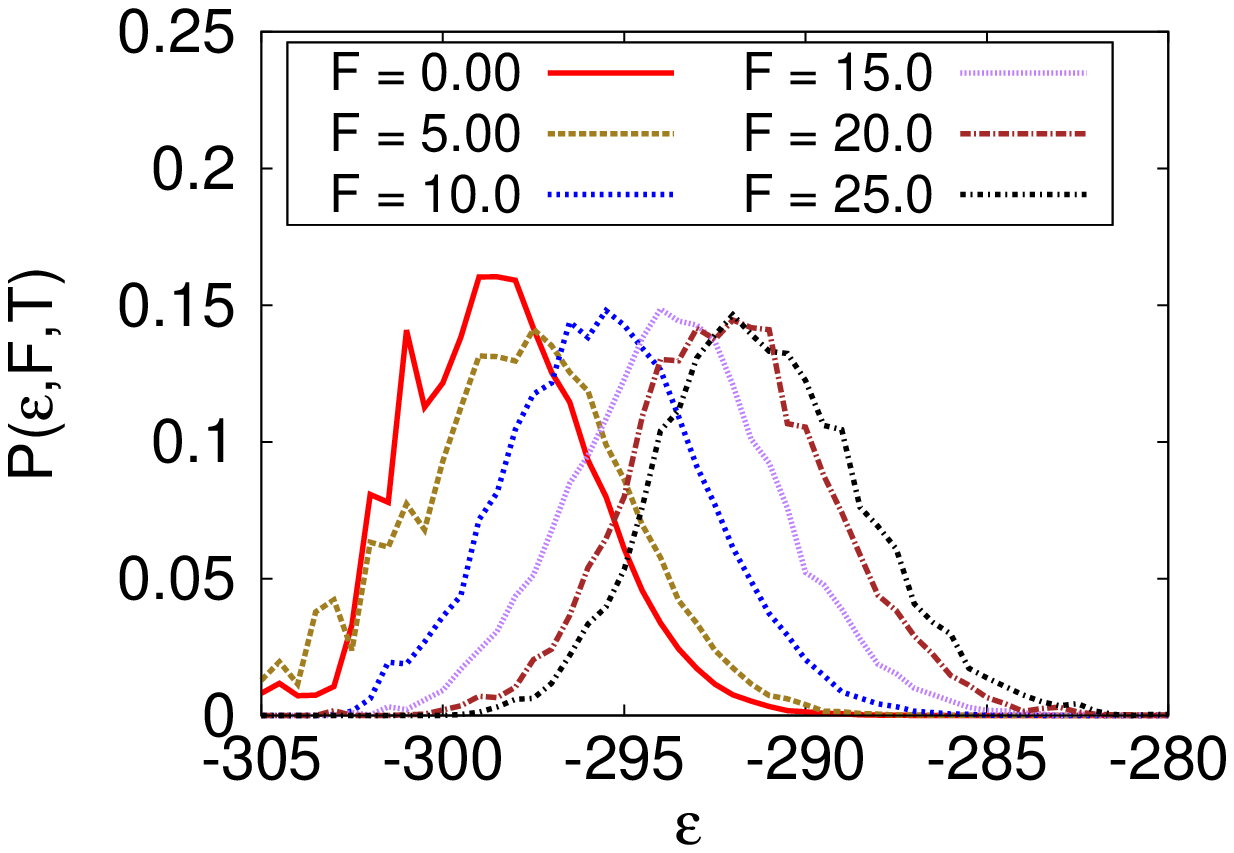}
\end{minipage}

\begin{minipage}{0.45\textwidth}
\begin{flushleft}
 \textbf{(c)}
\end{flushleft}
 \includegraphics[width=\textwidth]{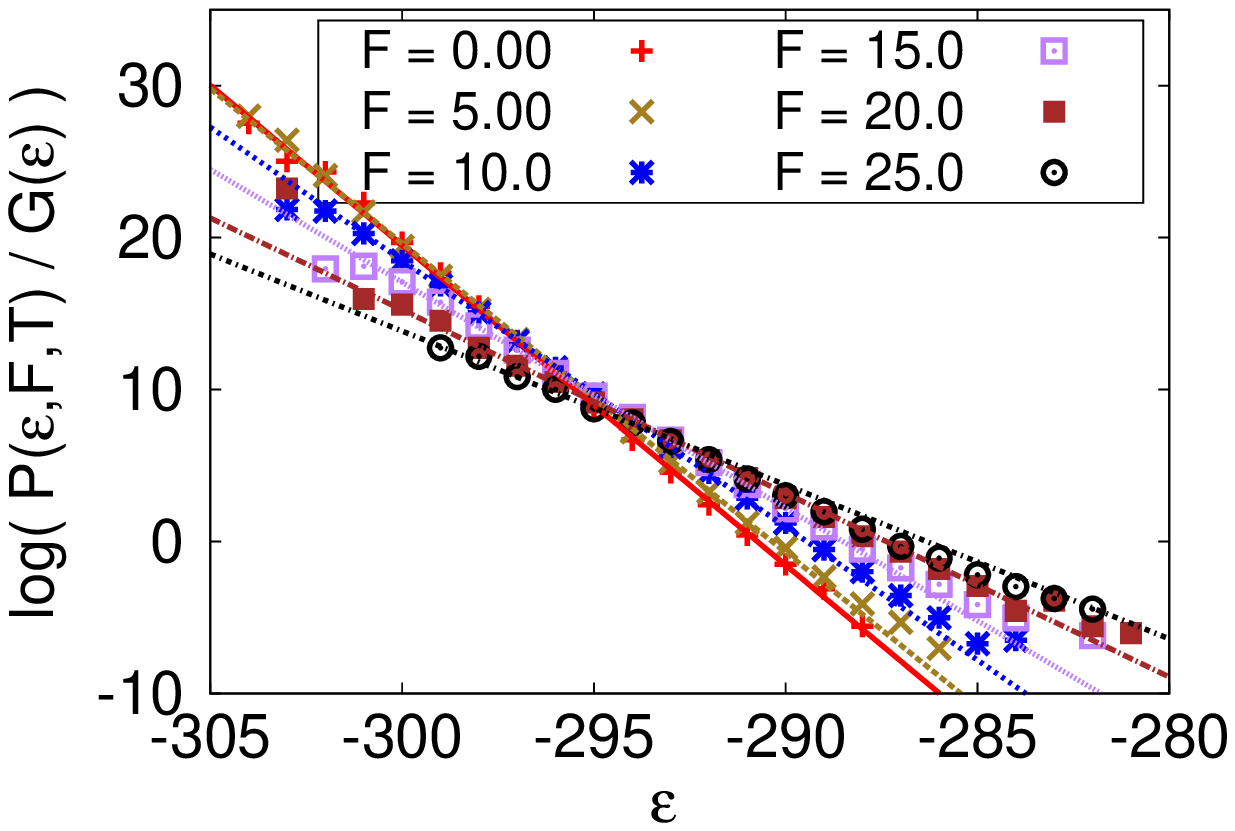}
\end{minipage} 
\begin{minipage}{0.45\textwidth}
\begin{flushleft}
 \textbf{(d)}
\end{flushleft}
 \includegraphics[width=\textwidth]{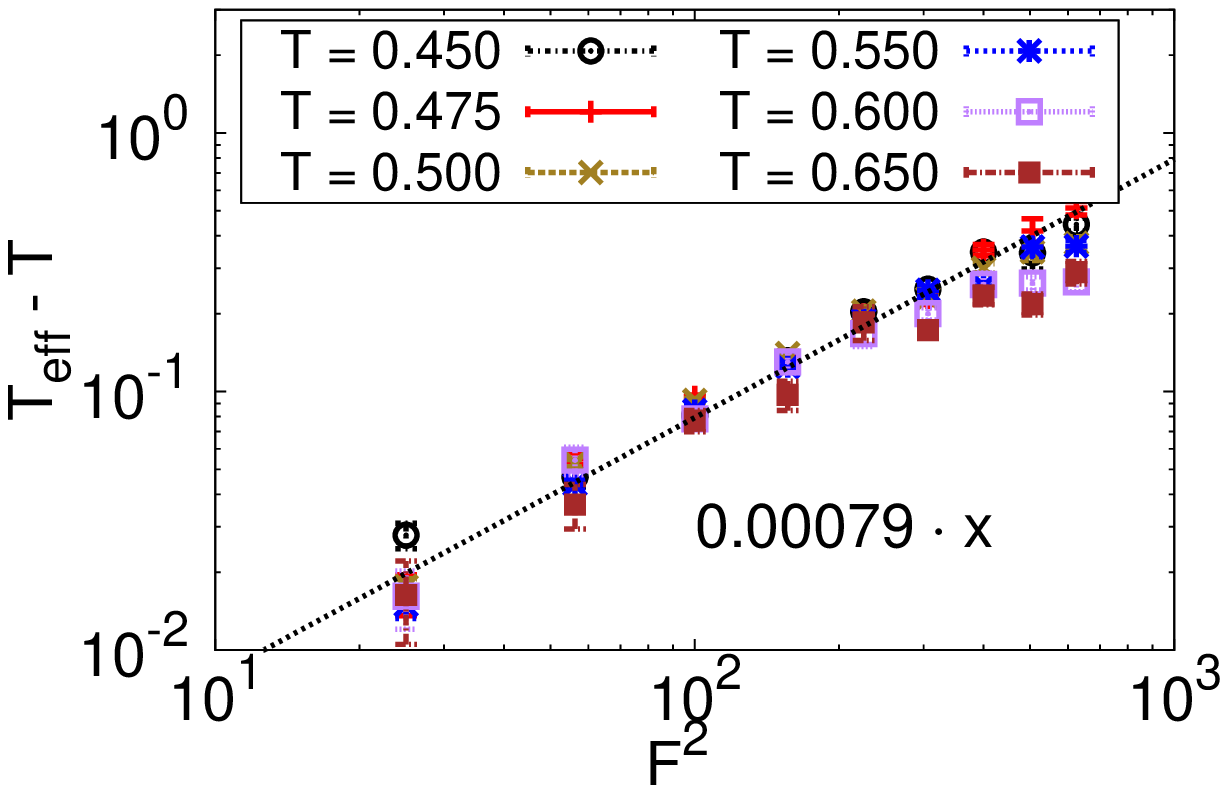}
\end{minipage}
\caption{Distribution of MB energies. (a) Probability distribution $P(\epsilon,T)$ of MB energies $\epsilon$ of quiescent systems at different temperatures $T$. (b) Probability distribution $P(\epsilon,F,T)$ of the microrheologically driven system at $T=0.475$ for different forces $F$. (c) Natural logarithm of the energy distribution $P(\epsilon,F,T)$ divided by MB density of states $G(\epsilon)$. The lines represent linear fits to the data points via \eq{eq:defTeff}. (d) Effective temperature $T_\eff$ of microrheologically driven systems for different forces and temperatures, rescaled in the spirit of \eq{eq:TeffF}. The black dotted line corresponds to a linear fit to the data points to determine the parameter $\chi$ in \eq{eq:TeffF}.}
\label{fig:energy}
\end{figure*}

The thermodynamic state of the system is characterized by the Boltzmann distribution of energies $P(\epsilon,T)$. Numerically, $P(\epsilon,T)$ can simply determined by computing the energy of a MB at arbitrary chosen points in time. In \fig{fig:energy}{(a)}, $P(\epsilon,T)$ is shown for a quiescent system at different temperatures. All curves display a Gaussian shape, which is shifted to larger values as the temperature is increased. In contrast to the average value, the width as well as the height of the distribution is not dependent on the temperature.

In the quiescent system for $T\lesssim0.7$, at which the anharmonic contributions of the local minimum are fully negligible \cite{Buechner1999}, the energy distribution results from the density of states (DOS) $G(\epsilon)$ via

\begin{equation}
 P(\epsilon,T) \propto G(\epsilon)~e^{-\epsilon/T},
 \label{eq:PepsT}
\end{equation}
where the states are weighted by the thermal energy of the system. With this equation, one can determine the underlying DOS by rescaling the distributions shown in \fig{fig:energy}{(a)}. As it is shown in \cite{Heuer2000}, the underlying DOS is a Gaussian with the same width as $P(\epsilon,T)$.

The dynamics of the system can still be described for $F>0$ in terms of the original ($F=0$) potential energy minima since the minima including the external force are, as we show in the Supplemental Material \cite{supp}, basically the same in the range of forces displayed in \fig{fig:energy}. It will be of key importance how the additional force changes the kinetic and thermodynamic properties of the system. In \fig{fig:energy}{(b)} $P(\epsilon,F,T)$ is shown for different force strengths. One observes that the Gaussian shape of the distribution is recovered also in the non-equilibrium case, but the distribution is shifted to higher energies. The width of the distribution is not altered by the application of the force. This is remarkable, since it displays the same characteristics which have already been observed for $P(\epsilon,T)$ in the quiescent system at different temperatures, see \fig{fig:energy}{(a)}. In particular, microrheological forcing and heating are indistinguishable in terms of the energy distribution.

This striking similarity between forcing and heating motivates the definition of an effective temperature which characterizes the shift of $P(\epsilon,F,T)$. Since, as mentioned above, $G(\epsilon)$ is by construction independent from the force, we can rewrite \eq{eq:PepsT} and insert the $P(\epsilon,F,T)$ for the energy distribution: 

\begin{equation}
  \epsilon/T_\eff = -\log\left(\frac{P(\epsilon,F,T)}{G(\epsilon)}\right) + \text{const}.
 \label{eq:defTeff}
\end{equation}
Thereby it is implicitly assumed that the energy distribution is a Boltzmann distribution, however, controlled by the effective temperature $T_\eff$ rather than the bath temperature $T$. The corresponding data are shown in \fig{fig:energy}{(c)} for different forces. It can be observed that plotting the data according to \eq{eq:defTeff} offers well defined straight lines whose slopes can be interpreted as the inverse effective temperatures. This observation supports the idea that the energy distribution can be regarded at least formally as Boltzmann distributed.

The definition for $T_\eff$ used above is similar to that in~\cite{Kob2000,Sciortino2001,Lacks2004} where it has been applied to characterize the non-equilibrium thermodynamics of a aging system and a system under an oscillatory shear strain. However, in these references only the average value of the energy distribution has been used to determine $T_\eff$. While in this particular case both definitions would lead essentially to the same values, it has been found in another context~\cite{Diezemann2011}, that the average value alone might be misleading, since it can result from very different distributions. Using the procedure depicted above would uncover such deviations by not showing straight lines in \fig{fig:energy}{(c)}. For later purposes, we rewrite \eq{eq:PepsT} for $F>0$ according to the effective temperature defined in \eq{eq:defTeff}:

\begin{equation}
 \frac{P(\epsilon,F,T)}{G(\epsilon)} \propto e^{-\epsilon/T_\eff}.
 \label{eq:PepsF}
\end{equation}

How is this effective temperature dependent on the external force? Since the shift of the energy distribution must be symmetric concerning the direction of the applied force, it reasonable to assume that it is an even function in $F$. Furthermore, in the limit of $F\to0$, $T_\eff$ must recover the bath temperature $T$. Because of this, a natural guess for the force dependence of $T_\eff$ is

\begin{equation}
 T_\eff = T + \chi F^2
\label{eq:TeffF}
\end{equation}
where $\chi$ is a force and temperature independent parameter. A similar guess has been justified by a mean-field theory of a Brownian particle in a viscoelastic medium \cite{Santamaria-Holek2011}. In this reference, the prefactor has, however, been identified with the mobility of the Brownian particle, which in the present case is strongly dependent on force and temperature. Thus, the relevance of the results from \cite{Santamaria-Holek2011} for microrheology of supercooled liquids is unclear.

Equation \ref{eq:TeffF} is tested in \fig{fig:energy}{(d)} for the microrheologically driven systems at different temperatures. It turns out that \eq{eq:TeffF} allows to superimpose the thermodynamic data points for a very broad range of forces onto a master curve. Only at high temperatures and large forces, small deviations occur. By fitting the master curve with a linear function, on can determine the value of $\chi$ for this system as $\chi=0.00079$. An estimate for the value of $\chi$ based on properties of the underlying PEL will be given at the end of the next subsection.

Instead of \eq{eq:TeffF}, one could also think about a scaling like 

\begin{equation}
 \beta_\eff = \beta-\chi(\beta F)^2,
 \label{eq:BeffF}
\end{equation}
with $\beta_\eff$ denoting the inverse effective temperature and $\beta$ denoting the inverse bath temperature, respectively. \eq{eq:BeffF} would fulfill the symmetry conditions discussed above as well and in the limit of $F\to0$, both \eq{eq:TeffF} and \eq{eq:BeffF} are identical. However, \eq{eq:BeffF} offers a much poorer scaling for large forces, whereas \eq{eq:TeffF}, as already discussed, allows a nearly perfect scaling for all forces and temperatures (see \fig{fig:energy}{(d)}).

\subsection{Escape kinetics}

\begin{figure*}
\begin{minipage}{0.45\textwidth}
\begin{flushleft}
 \textbf{(a)}
\end{flushleft}
\includegraphics[width=\textwidth]{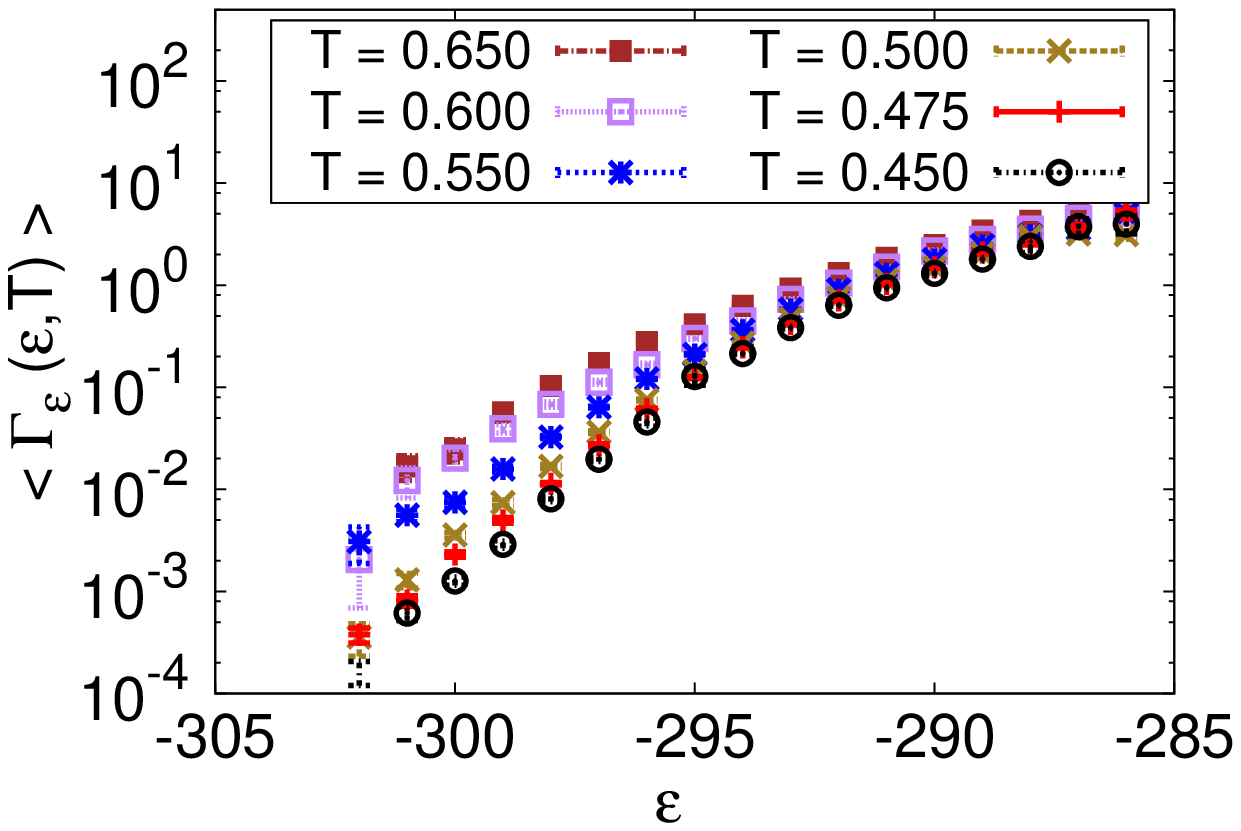}
\end{minipage}
\begin{minipage}{0.45\textwidth}
\begin{flushleft}
 \textbf{(b)}
\end{flushleft}
\includegraphics[width=\textwidth]{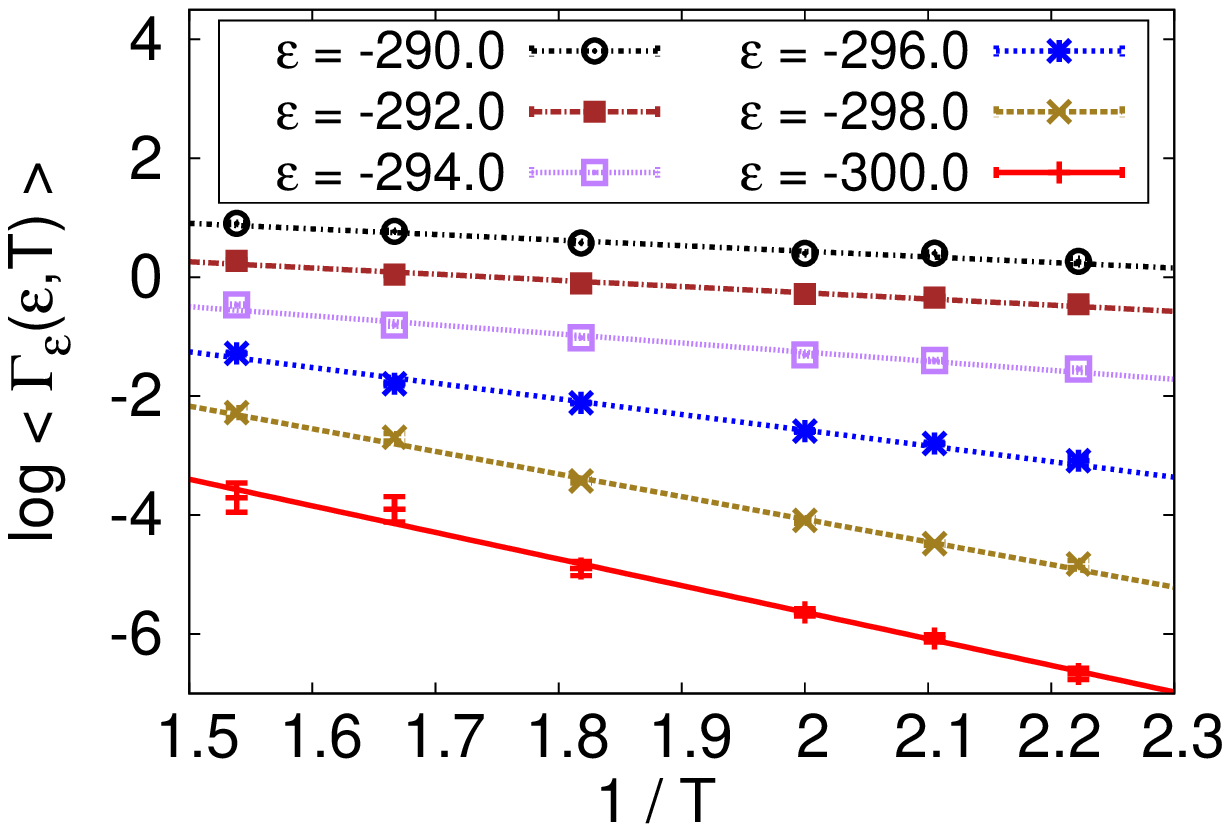}
\end{minipage}
\begin{minipage}{0.45\textwidth}
\begin{flushleft}
 \textbf{(c)}
\end{flushleft}
\includegraphics[width=\textwidth]{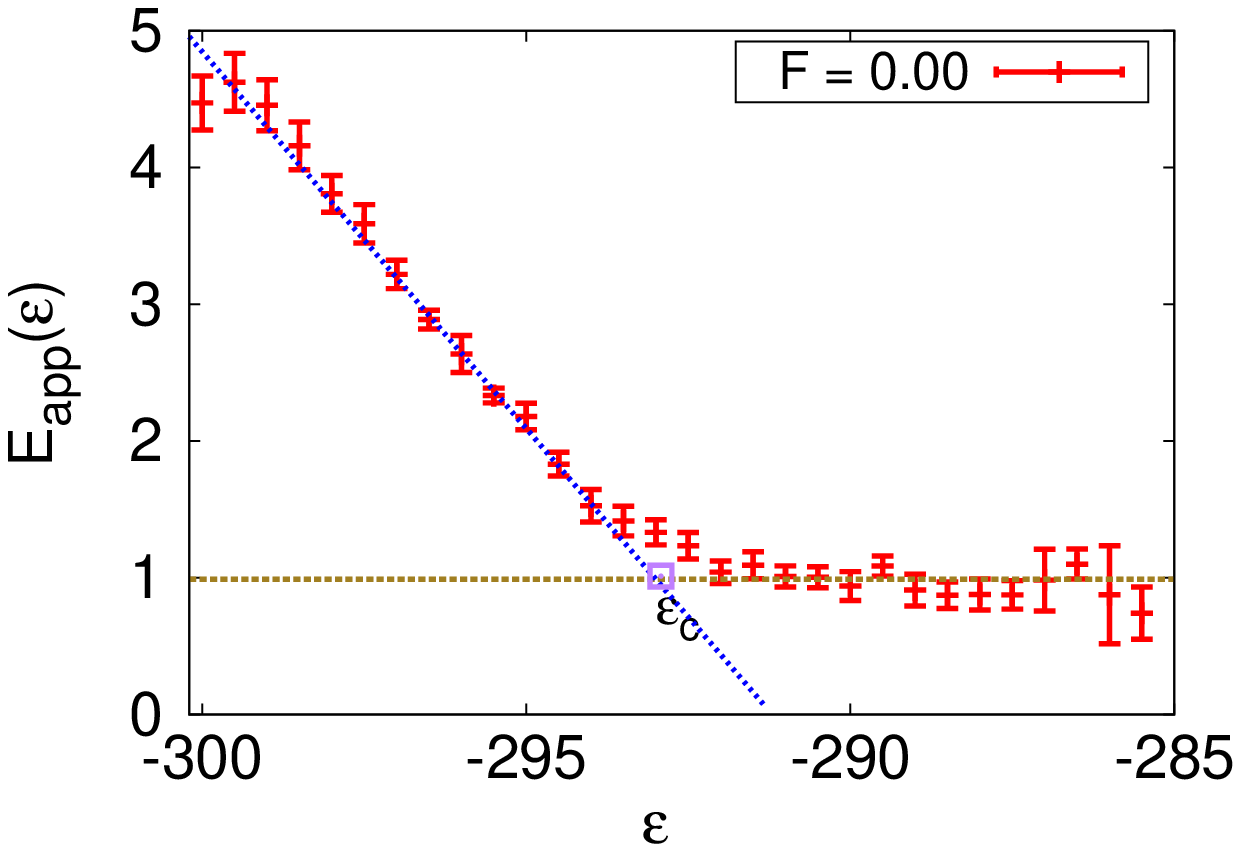}
\end{minipage}
\begin{minipage}{0.45\textwidth}
\begin{flushleft}
 \textbf{(d)}
\end{flushleft}
 \includegraphics[width=\textwidth]{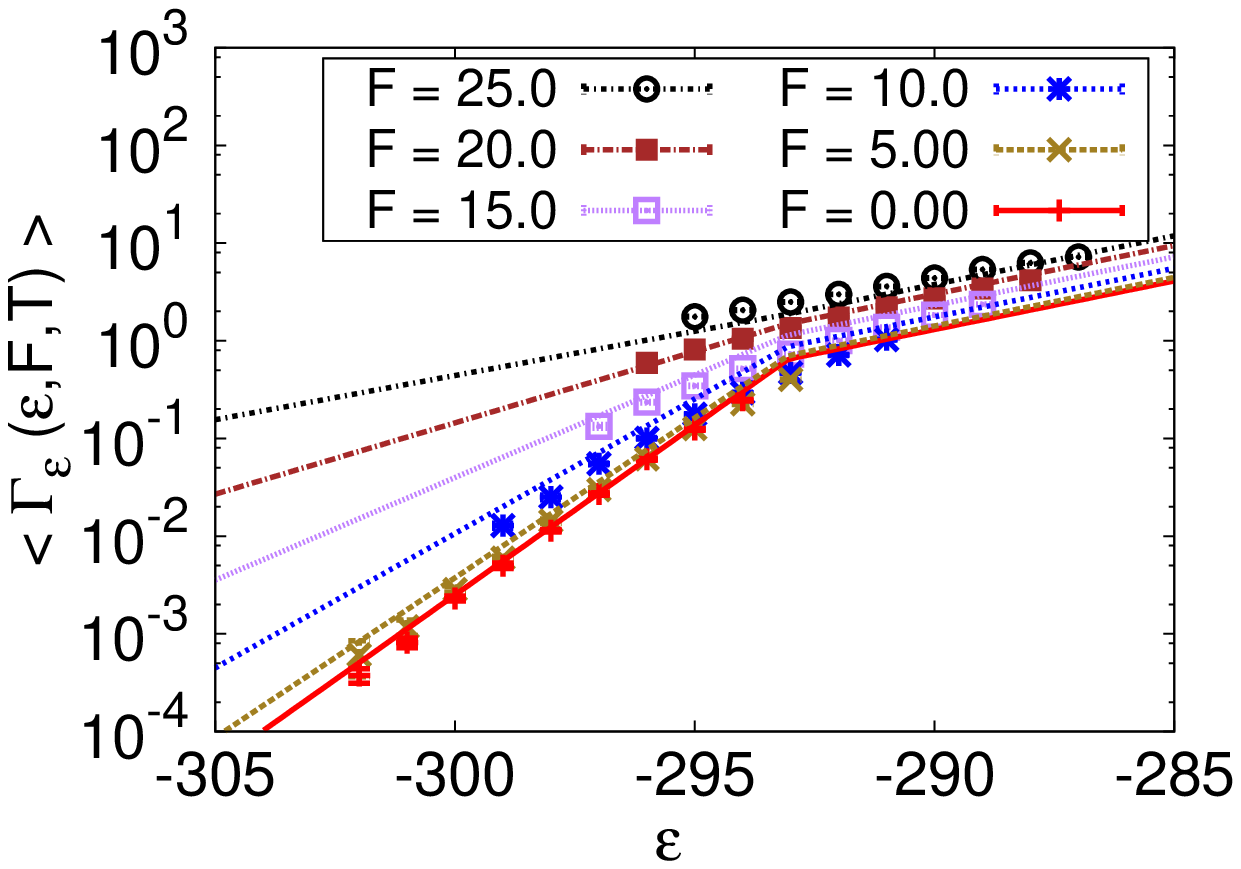}
\end{minipage}
\caption{Average rates to escape from a single MB. (a) Average escape rate $\av{\Gamma_\epsilon(\epsilon,T)}$ to escape from a MB of energy $\epsilon$ for quiescent systems at different temperatures $T$. (b) Natural logarithm of the average escape rates $\av{\Gamma_\epsilon(\epsilon,T)}$ for different MB energies $\epsilon$ as a function of the inverse temperature $1/T$. The lines correspond to linear fits to the data points via \eq{eq:GeT}. (c) Activation energies $E_\app(\epsilon)$ as a function of the energies of the MB. The single purple point marks the crossover energy $\epsc$, the lines are linear fits to the data points via \eq{eq:Eapp}. (d) Average escape rates $\av{\Gamma_\epsilon(\epsilon,F,T)}$ of the microrheologically driven system at $T=0.475$ and different forces $F$. The lines corresponds to the theoretical prediction via \eq{eq:GeF} where the the effective temperature has been determined via \eq{eq:TeffF}.}
\label{fig:kinetics}
\end{figure*}

The other central quantity is the average escape rate $\av{\Gamma_\epsilon(\epsilon,T)}$ to leave a MB of energy $\epsilon$. For a quiescent system at different temperatures, these local escape rates are shown in \fig{fig:kinetics}{(a)}. One finds that the temperature-dependence of the apparent escape rates is dependent on the energy of the MB. This result is already known from previous simulations~\cite{Doliwa2003b} and can be understood by the fact that the escape from singe MB, although they occur via a multi-minima path, can be efficiently described as a single activated process:

\begin{equation}
 \av{\Gamma_\epsilon(\epsilon,T)} = \Gamma_0(\epsilon) e^{-\beta E_\app(\epsilon)}.
 \label{eq:GeT}
\end{equation}
Thereby both, the apparent activation energy $E_\app(\epsilon)$, the system has to overcome, and the prefactor $\Gamma_0(\epsilon)$, which is of entropic nature and accounts for the number of possible paths to leave the minimum, are dependent on the energy of the particular MB. By fitting the escape rates of particular energies at different temperatures via \eq{eq:GeT}, one is able to determine $E_\app(\epsilon)$ and $\Gamma_0(\epsilon)$, see \fig{fig:kinetics}{(b)}. The resulting curve for $E_\app(\epsilon)$ is shown in \fig{fig:kinetics}{(c)}. One observes two distinct regimes:

\begin{equation}
 E_\mathrm{app}(\epsilon) = 
 \begin{cases}
  m(\epsc - \epsilon)+V_0 & \text{for } \epsilon<\epsc \\
  V_0 &\text{for } \epsilon>\epsc\\
  \end{cases} 
 \label{eq:Eapp}
\end{equation}
with the critical MB energy of the system $\epsc\approx-293$ and a slope $m\approx 0.55$. Since the dynamics of the system are merely determined by the slow escape rates, corresponding to large energy barriers, the following analysis will focus on the low-energy part of \eq{eq:Eapp}.

For the case $m=1$, the activation energy is in agreement with the trap model~\cite{Monthus1996}, where the MBs serve as traps: The system has to overcome a barrier equal to the depth of the trap plus a small additional barrier $V_0$. What is the meaning of the factor $m$ at low energies? Although our simulations are conducted with a very small system, only a small set of particles is moving at the same time~\cite{Vogel2004}. To estimate the implications for $E_\app(\epsilon)$ one may consider that the simulated system consists of $M$ non-interacting subsystems, each one relaxing independently from each other. The escape rates one measures is therefore the rate one observes for the overall system, i.e. sum of the rates of each individual subsystem $i$: $\Gamma(\epsilon=\sum_1^M \epsilon_i) = \sum_1^M \Gamma_i(\epsilon_i)$. If one now assumes that the energy of the system is distributed equally among the $M$ subsystems, one immediately obtains $\Gamma(\epsilon)=M\Gamma_i(\epsilon/M)$, thus, the actual barrier, which has to be overcome, is reduced by a factor $m=1/M$~\cite{Heuer2005}. These ideas give rise to the notion of an "extended trap model"~\cite{Heuer2005}. Considering that the activation energy is proportional to the trap depth (\eq{eq:Eapp}), one can rewrite \eq{eq:GeT} as

\begin{equation}
 \av{\Gamma_\epsilon(\epsilon,T)}^{-1/m} \propto e^{-\epsilon/T} \propto \frac{P(\epsilon,T)}{G(\epsilon)},
 \label{eq:GeTrap}
\end{equation}
where the right hand side has been identified with the ratio of the energy distribution and the density of states, see \eq{eq:PepsT}. This connection of the escape dynamics on the one side and the thermodynamics on the other side is a fundamental property of the extended trap model.

The escape rates of a microrheologically driven system are shown in \fig{fig:kinetics}{(d)} for a broad range of forces. One observes that the escape rates at low MB energies are strongly enhanced as compared to the quiescent case whereas at high MB energies the rates only slightly grow. By visual inspection on finds that this behavior is rather similar to what one observes in \fig{fig:kinetics}{(a)} for a quiescent system at different temperatures. This coincidence suggests that not only the energy distribution but also the escape kinetics are driven by an effective temperature.

This can further be justified in terms of the extended trap model: In \eq{eq:GeTrap}, a fundamental connection between the thermodynamics and kinetics is established which \emph{a priori} holds for the quiescent system only. For the driven system, the right hand side of this equation can be replaced by \eq{eq:PepsF}, thus, the control parameter of the thermodynamics is the effective temperature $T_\eff$. If \eq{eq:GeTrap} still holds for this system, it follows that also the kinetics are determined by $T_\eff$ defined in \eq{eq:TeffF}, i.e.

\begin{equation}
 \av{\Gamma_\epsilon(\epsilon,F,T)} =  \av{\Gamma_\epsilon(\epsilon,0,T_\eff)}.
 \label{eq:GeF}
\end{equation}
For this relation it is implicitly assumed that the factor $m$ of the extended trap model is not altered by the external force, thus, that the number of subsystems is the same as in the quiescent system. This assumption can be justified by the observation that the average number of particles, involved in a MB transition, is basically $F$-independent (see Supplemental Material \cite{supp} for more details). Thus, the value of $m$ defined in \eq{eq:Eapp}, describes the $F>0$ case as well.

Equation (\ref{eq:GeF}) can be tested by inserting equations (\ref{eq:TeffF}), (\ref{eq:GeT}) and (\ref{eq:Eapp}) for different forces, as it is shown in \fig{fig:kinetics}{(d)}. One finds a very good agreement between the theoretical prediction and the numerical data. Our results show that both thermodynamics and kinetics are controlled by the same effective temperature $T_\eff$ which is dependent on a single parameter $\chi$.

Its microscopic origin can be qualitatively explored in terms of a one-dimensional cosine potential which is tilted along the force direction. As it is demonstrated in the Supplemental Material \cite{supp}, one expects for this toy model that the escape rate out of a single minimum is for small forces given by $\exp(-E/T_\eff)$ with $T_\eff = T+a^2/(8 E) F^2$ where $E$ denotes the barrier and $a$ the distance between two minima. For the binary Lennard-Jones mixture, a reasonable guess for the energy scale is the width $\sigma$ of the DOS or, equivalently, of the energy distribution $P(\epsilon,T)$. The distance between to MB is given by the squared diffusive length $\av{a^2}=\lim_{n\to\infty}\av{x^2(n)}/n$ of a single particle which can be determined from the mean-squared displacement $\av{x^2(n)}$ of the particle after a number of MB transitions $n$. Since both, $\sigma$ and $\av{a^2}$ are independent from temperature \cite{Doliwa2003c}, also the value of $\chi_\text{est} = \av{a^2}/(8 \sigma)$ does not dependent on $T$. With $\av{a^2} = 0.00351$ and $\sigma = 2.65$, one can estimate $\chi_\text{est} = 0.00017$ which agrees with the actually measured value of $\chi=0.00079$ up to a factor $4.6$. Such deviations can be rationalized by considering that barriers between two MB do not lie exactly in the middle of two MB but are shifted closer to the starting minimum. In any event, already this simple model offers a reasonable estimate for the parameter $\chi$, which is purely based on properties of the quiescent energy landscape. 

\subsection{Force-dependent mobility}

\begin{figure*}
\begin{minipage}{0.45\textwidth}
\begin{flushleft}
 \textbf{(a)}
\end{flushleft}
\includegraphics[width=\textwidth]{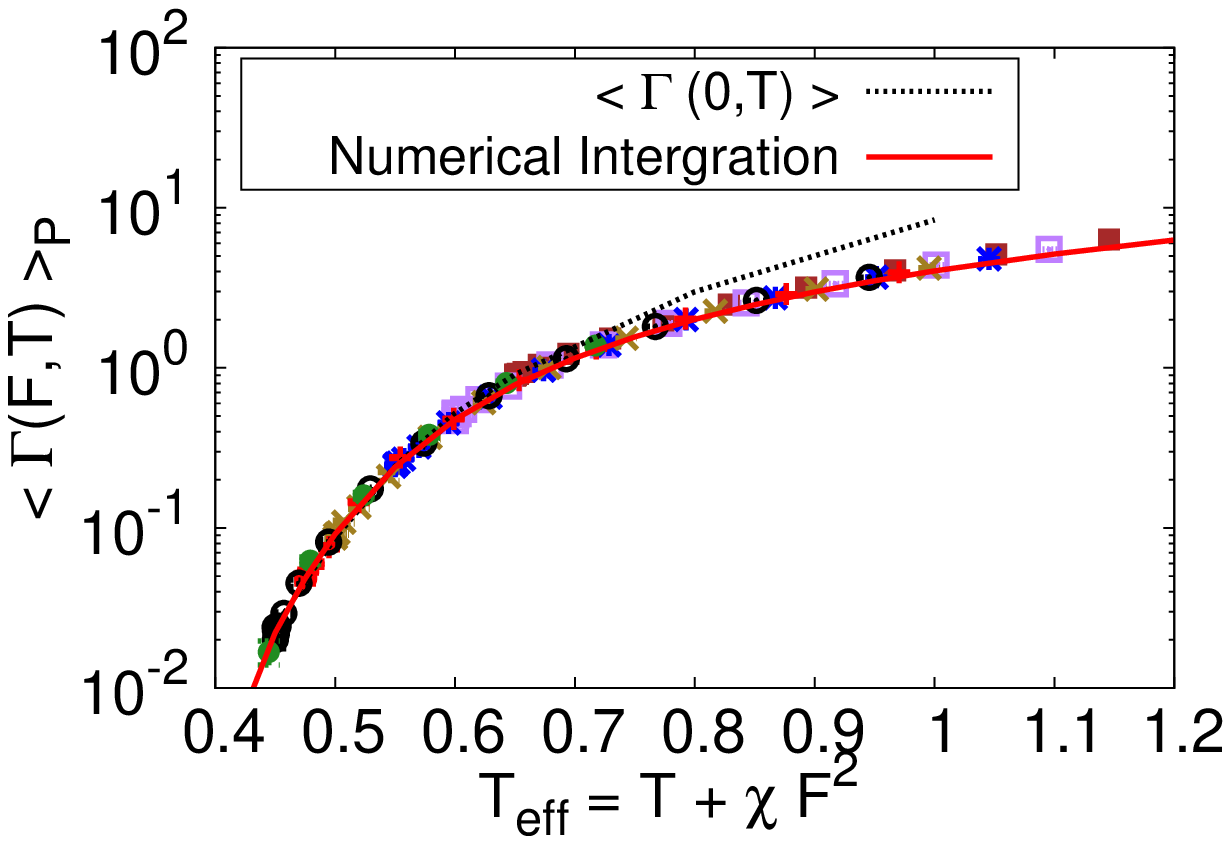}
\end{minipage}
\begin{minipage}{0.45\textwidth}
\begin{flushleft}
 \textbf{(b)}
\end{flushleft}
 \includegraphics[width=\textwidth]{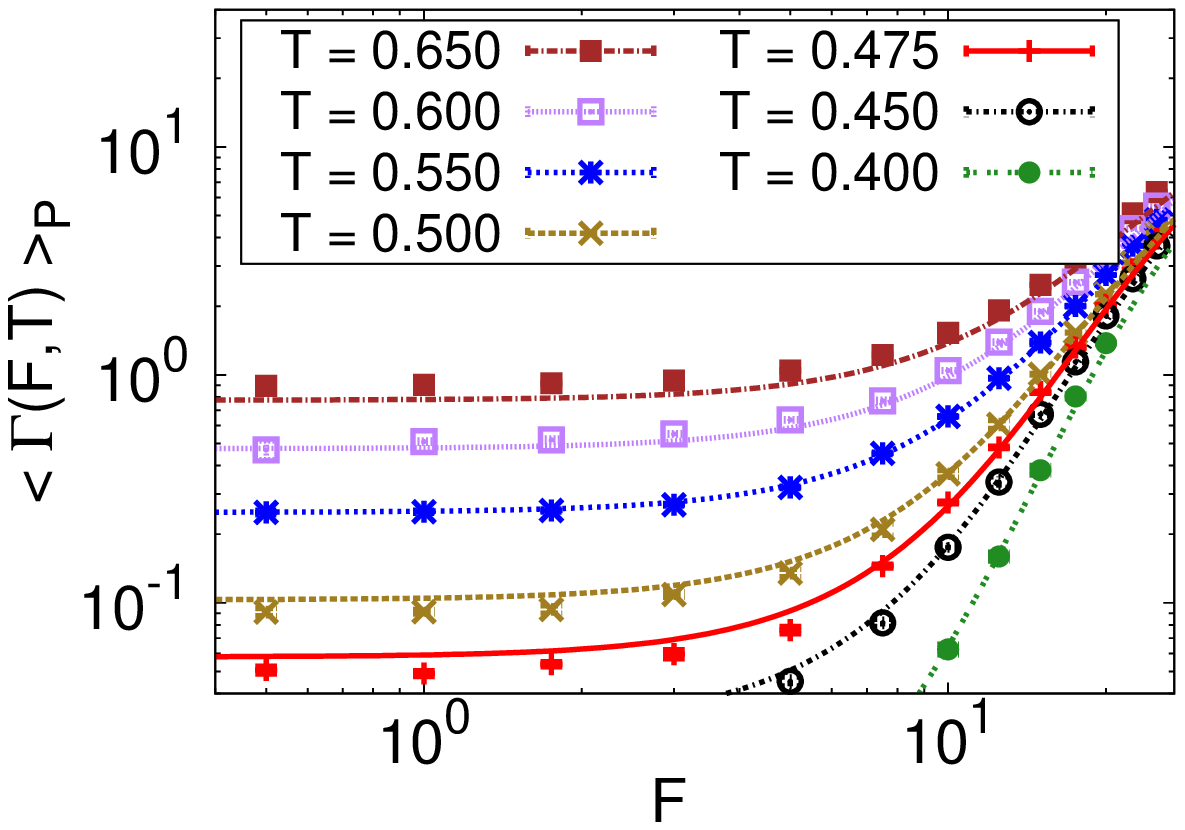}
\end{minipage} 

\begin{minipage}{0.45\textwidth}
\begin{flushleft}
 \textbf{(c)}
\end{flushleft}
 \includegraphics[width=\textwidth]{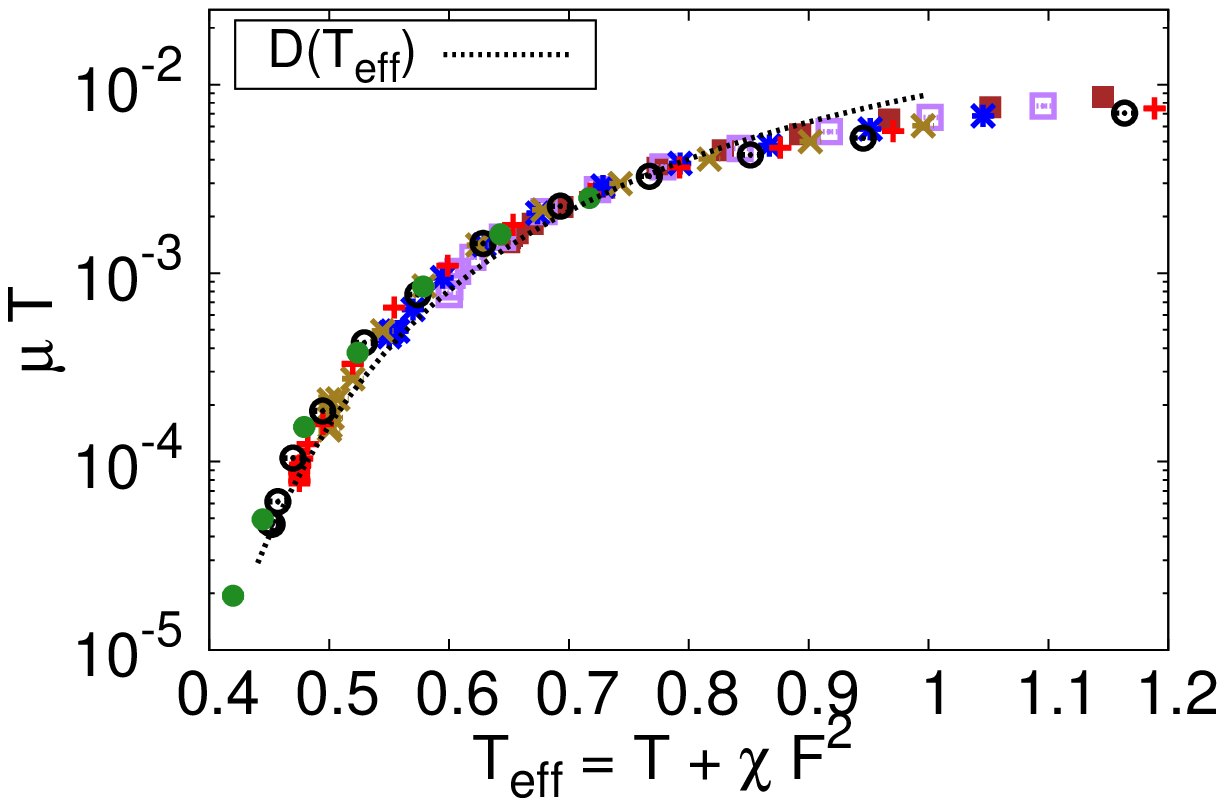}
\end{minipage}
\begin{minipage}{0.45\textwidth}
\begin{flushleft}
 \textbf{(d)}
\end{flushleft}
 \includegraphics[width=\textwidth]{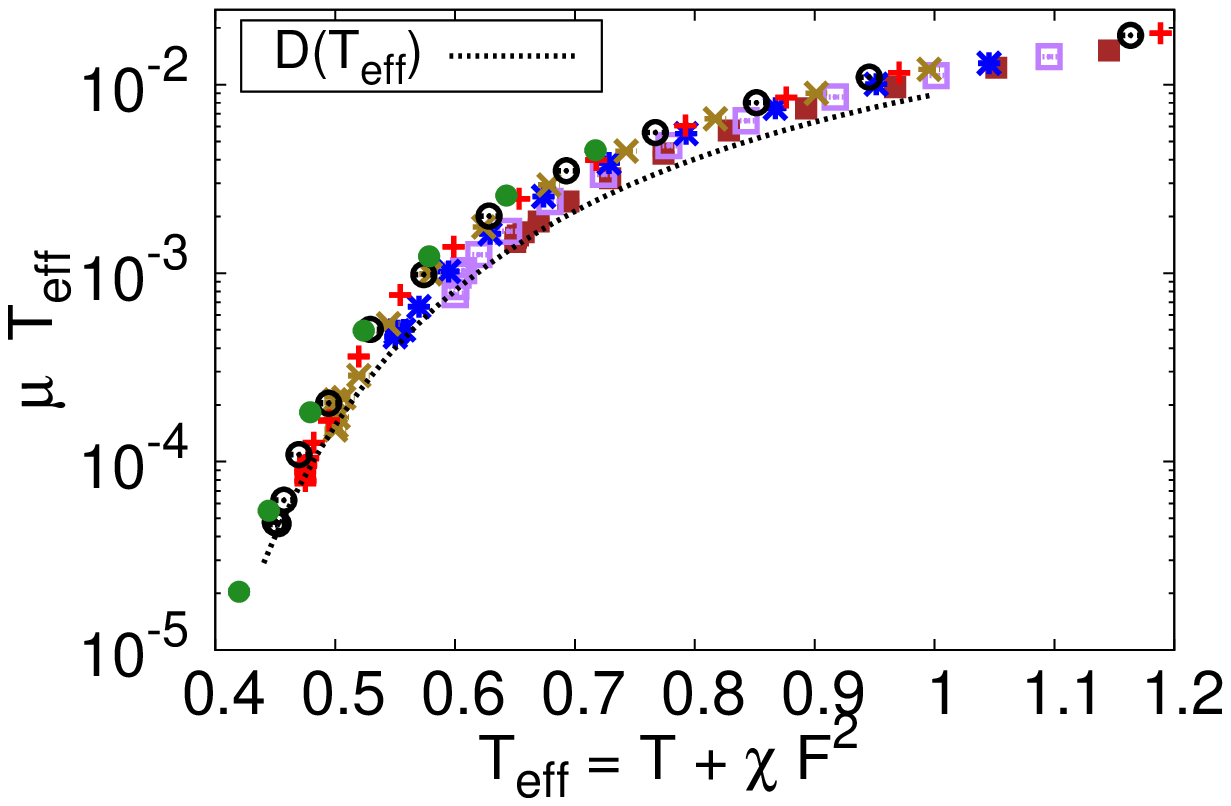}
\end{minipage}
\caption{Force-Temperature Superposition of the nonlinear mobility. (a) Average escape rate $\av{\Gamma(F,T)}$ for different forces $F$ and bath temperatures $T$ as a function of the effective temperature $T_\eff = T+\chi\cdot F^2$. The color code for the different bath temperatures is the same as in (b). The black dotted line corresponds to the average escape time $\av{\Gamma(0,T)}$, determined from MD simulations of a quiescent system, the red solid line to the theoretical prediction by evaluating \eq{eq:defAvGamma}. (b) Average escape rate $\av{\Gamma(F,T)}$ as a function of the force $F$ at different bath temperatures $T$. The lines corresponds to the theoretical prediction via \eq{eq:defAvGamma}. Please note that the data points for $T=0.4$ (full green circles) correspond to a bath temperature below the mode-coupling temperature $\Tc = 0.435$~~\cite{Kob1995}. (c) Mobility $\mu$ times the bath temperature as a function of the effective temperature $T_\eff$. The color code of the data points is the same as in (b). The black dotted line corresponds to the theoretical prediction according to \eq{eq:MuF2}. (d) The mobility $\mu$ times the effective temperature $T_\eff$ as a function of the effective temperature $T_\eff$. The color code of the data points is the same as in (b). The black dotted line is the same as in (c).}
\label{fig:FTSmu}
\end{figure*}

In the next step, one can revisit \eq{eq:defAvGamma} to study the force-dependence of the average escape rate $\av{\Gamma(F,T)}_P$ with provides the time scale of the mobility $\mu$. Since both of its constituents are characterized exclusively by the effective temperature, also $\av{\Gamma(F,T)}_P$ is a function of $T_\eff$ only. This result is shown in \fig{fig:FTSmu}{(a)} where the average escape rate of a broad range of forces and temperature is shown as a function of their effective temperatures. One finds that all data points can be collapsed onto a single master curve. This property may be denoted as \emph{Force-Temperature Superposition}.

A subtlety emerges due to the fact that \eq{eq:PepsT} has been used to compute the energy distribution for \eq{eq:defAvGamma}, since this equation requires anharmonic effects to be fully negligible ($T\lesssim0.7$). This difference is directly visible in \fig{fig:FTSmu}{(a)}. Interestingly, one finds that the prediction via \eqs{eq:defAvGamma} and (\ref{eq:PepsT}) offers a better description of the superimposed data points. The observation that in case of microrheology the harmonic approximation holds for large effective temperatures suggest that the fluctuations inside a minimum are still controlled by the low bath temperature given by the thermostat. This is is in agreement with the result for a sheared system in \cite{Berthier2002} where it has been found that the short-time dynamics are determined by the bath temperature whereas the dynamics on long timescales are governed by a higher effective temperature.

In \fig{fig:FTSmu}{(b)} the average escape rates are shown as a function of the force for different temperatures, together with the theoretical prediction based on the numerical integration of \eq{eq:defAvGamma}. One finds a quantitative agreement for all forces and temperatures. We would like to emphasize that the data for a temperature $T=0.4$ (full green circles in \fig{fig:FTSmu}{}) has been included which is below the mode-coupling temperature of the system ($\Tc=0.435$~~\cite{Kob1995}). For this temperature, the quiescent state is inaccessible due to very long relaxation times, thus, it can be regarded as below the (computer) glass transition. Since \eq{eq:defAvGamma} can still be evaluated, the present theory nevertheless allows a qualitative prediction of its highly non-equilibrium state.

The insights obtained above can eventually be used to predict the force-dependent mobility of the tracer particle in the supercooled liquid. For this purpose one has to recall, that the mobility in terms of the PEL approach given by 
\begin{equation}
\mu = x(F,T)~\av{\Gamma(0,T_\eff)}_P,
\label{eq:MuF1}
\end{equation}
 where $x(F,T)$ denotes the spatial part which depends on force and temperature. However, as it has been shown in \cite{Schroer2013} and the Supplemental Material \cite{supp}, the spatial part displays only a weak dependence on the external force and is essentially given by
 
 \begin{equation}
 x(F,T)\approx\av{a^2}/(2T)
 \label{eq:xF}
 \end{equation}
  where $\av{a^2}$ again denotes the diffusive length scale of a single particle in the quiescent system. The one-dimensional diffusion coefficient $D$ of a single particle in quiescent system is given by \cite{Doliwa2003c}
  
  \begin{equation}
   D(T) = \left(\av{a^2}/2\right)~\av{\Gamma(T)}_P\text{,}
    \label{eq:DT}
  \end{equation}

  thus, one can combine \eqs{eq:MuF1}-(\ref{eq:DT}) to obtain for the mobility:
  
 \begin{equation}
  \mu = D(T_\eff) / T.
  \label{eq:MuF2} 
 \end{equation}
Here, the force-dependent mobility is brought into relation with the temperature dependence of the equilibrium diffusion coefficient and can therefore be tested with real space quantities only.

The mobility $\mu$ as a function of the effective temperature is shown in \fig{fig:FTSmu}{(c)}. There one can observe that again all data points fall onto a single master curve which is in a good agreement with the theoretical predictions via \eq{eq:MuF2}. The deviations at rather low effective temperatures occur due to the fact, that the force-dependence of the spatial part in \eq{eq:MuF1} has not been considered in \eq{eq:MuF2}. At high effective temperatures, the theoretical prediction differs from the superimposed data point due to the breakdown of the harmonic approximation, similar to what already has been observed in \fig{fig:FTSmu}{(a)} for the average escape rate.

We would like to mention that the structure \eq{eq:MuF2} may seem to be a little bit odd, since two different temperatures explicitly enter: The effective temperature which determines the diffusion coefficient and the bath temperature in the denominator. However, this behavior is a direct result of the PEL approach, since in \eq{eq:xF}, $T$ cannot be replaced by $T_\eff$. As a consequence, the application of $\mu = D(T_\eff)/T_\eff$ for the rescaling of the data results in a much poorer Force-Temperature Superposition in the nonlinear regime, see \fig{fig:FTSmu}{(d)}. In summary, \eq{eq:MuF2} offers very reasonable description of the force-dependent mobility, based on real space properties of the quiescent system and a single parameter $\chi$ for a broad range of forces and temperatures.

\subsection{Diffusive properties}

\begin{figure*}
\begin{minipage}{0.45\textwidth}
\begin{flushleft}
 \textbf{(a)}
\end{flushleft}
\includegraphics[width=\textwidth]{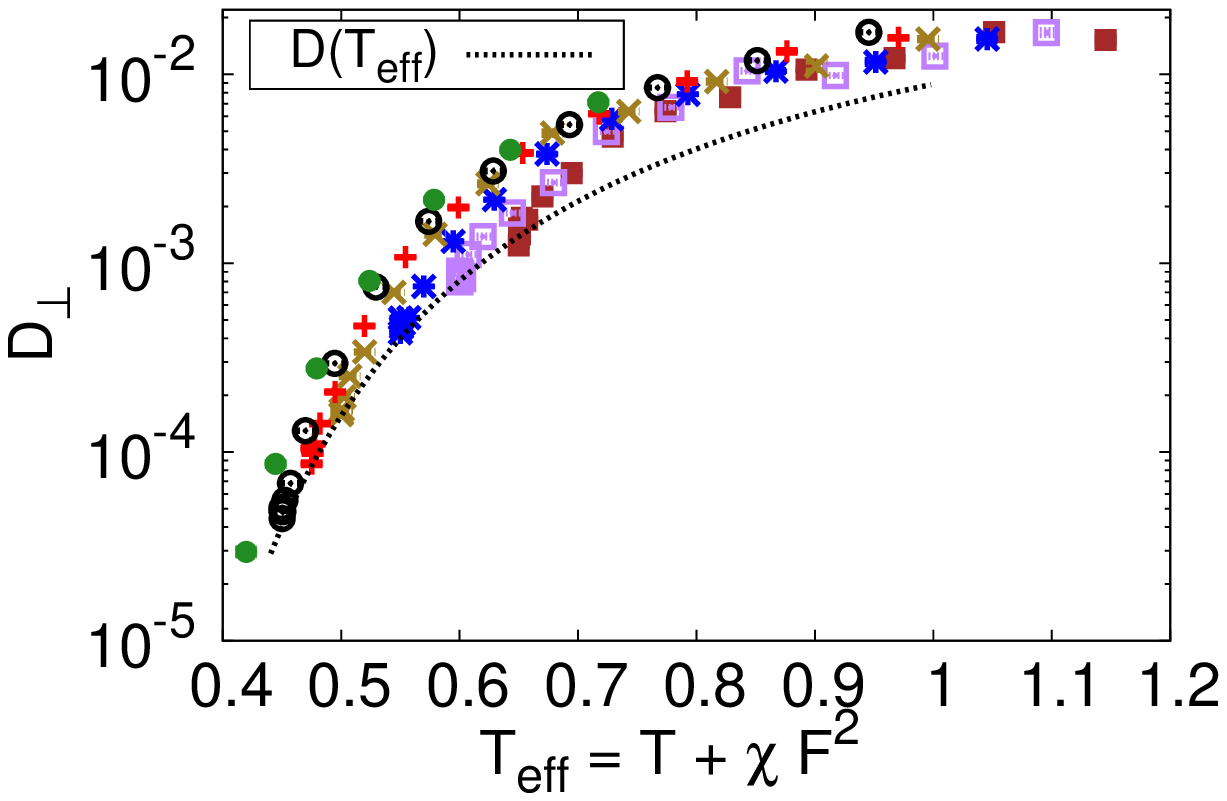}
\end{minipage}
\begin{minipage}{0.45\textwidth}
\begin{flushleft}
 \textbf{(b)}
\end{flushleft}
 \includegraphics[width=\textwidth]{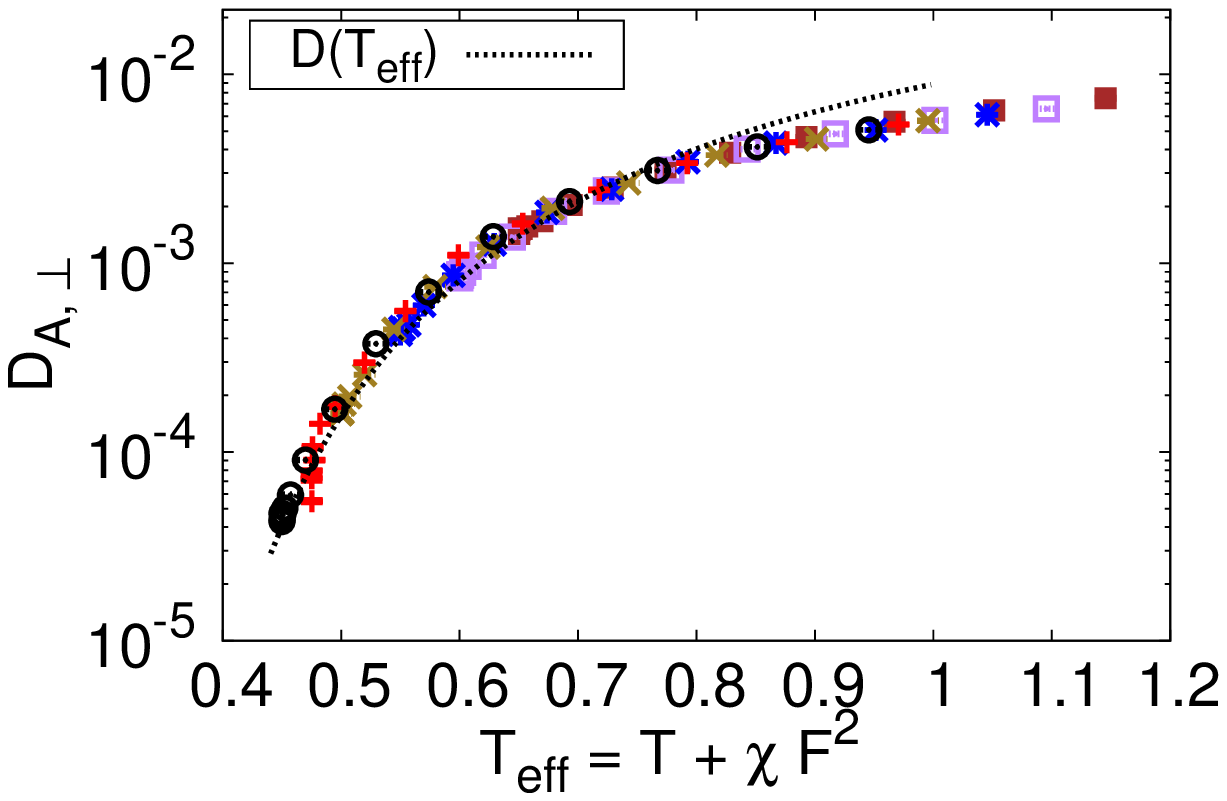}
\end{minipage} 
\begin{minipage}{0.45\textwidth}
\begin{flushleft}
 \textbf{(c)}
\end{flushleft}
 \includegraphics[width=\textwidth]{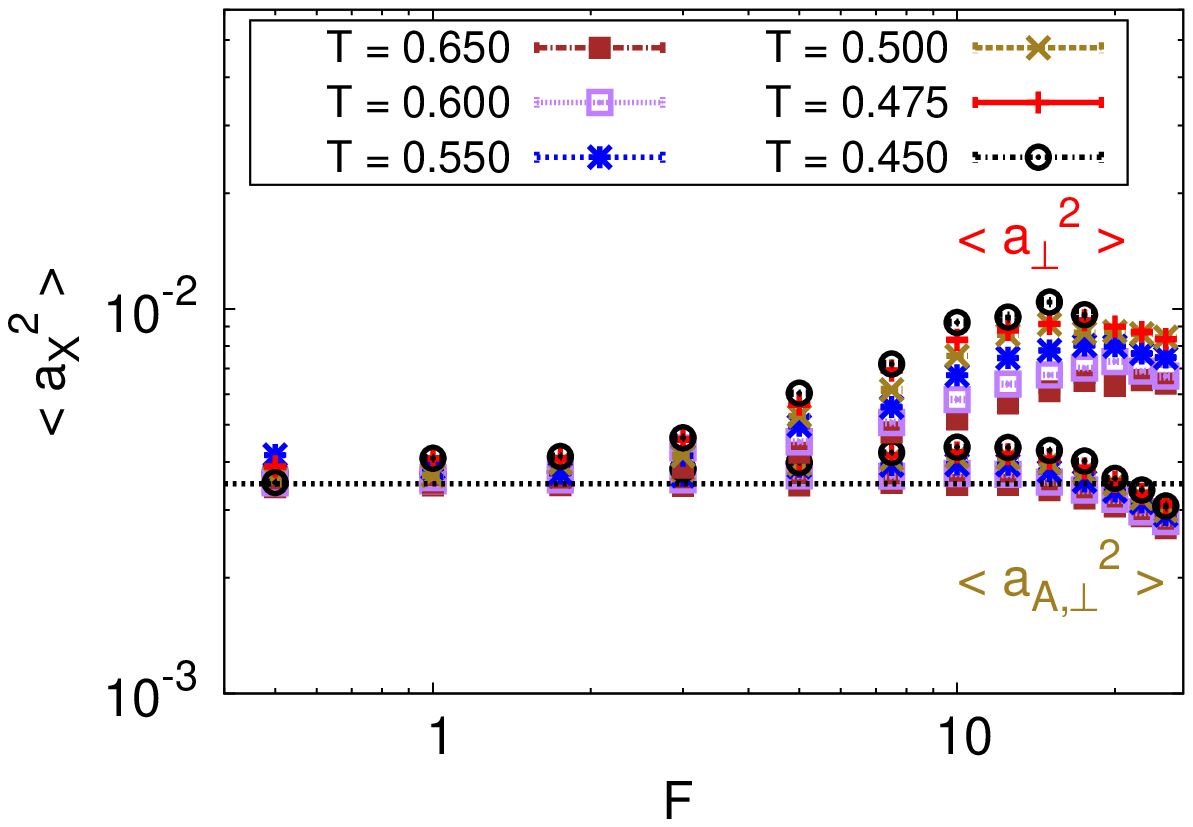}
\end{minipage}
\begin{minipage}{0.45\textwidth}
\begin{flushleft}
 \textbf{(c)}
\end{flushleft}
 \includegraphics[width=\textwidth]{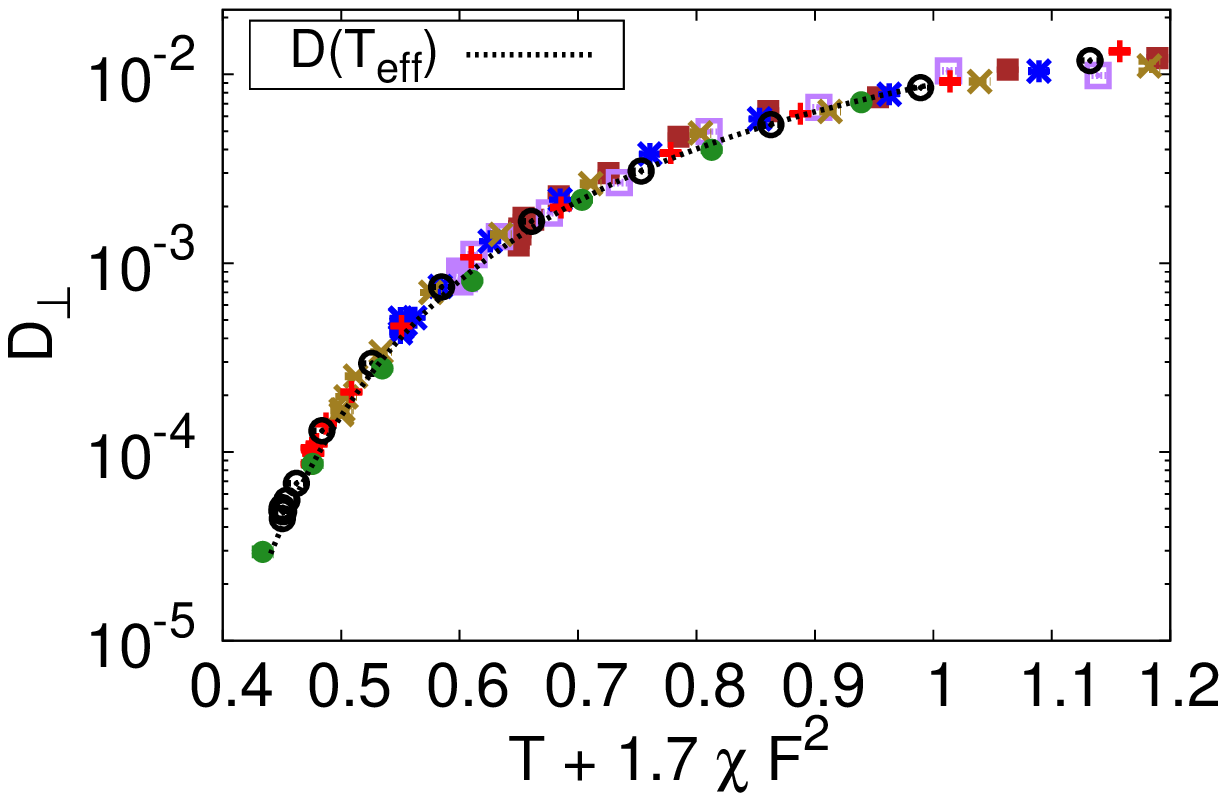}
\end{minipage}
\caption{Force-Temperature Superposition of the orthogonal diffusion coefficients.
(a) Orthogonal diffusion coefficient $D_\perp$ of the tracer particle as a function of the effective temperature $T_\eff = T+\chi\cdot F^2$. The color code is the same as in \fig{fig:FTSmu}{(b)}. The black dotted line corresponds to the theoretical prediction according to \eq{eq:DF2}. 
(b) Orthogonal diffusion coefficient $D_{\text{A},\perp}$ of the bath particles of type A as a function of the effective temperature $T_\eff = T+\chi\cdot F^2$. The color code is the same as in \fig{fig:FTSmu}{(b)}. The black dotted line corresponds to the theoretical prediction according \eq{eq:DF2}.
(c) Orthogonal diffusive length scales $\av{a^2_\text{X}}$ of the tracer and the type A bath particles for different forces $F$ and temperatures $T$. The upper branch ($\av{a^2_\perp}$) belongs to the diffusive length scale of the tracer particle, the lower branch ($\av{a^2_{\text{A},\perp}}$) to that of the bath particles of type A. The black dotted line corresponds to the temperature-independent \cite{Doliwa2003c} diffusive length scale of the quiescent system.
(d) Orthogonal diffusion coefficient $D_\perp$ of the tracer particle as a function of the scaled effective temperature $T+1.7~\chi F^2$. The color code of the data points is the same as in \fig{fig:FTSmu}{(b)}. The black dotted line is the same as in (a).}
\label{fig:FTSD}
\end{figure*}

Finally, one may ask whether the effective temperature also governs the diffusive properties of the system. For this purpose, we have determined the nonlinear diffusion coefficient $D_\perp$ of the tracer particle orthogonal to the force direction as well as $D_{A,\perp}$ of the type A bath particles in the same direction by fitting the long-time limit of the mean-squared displacements. The orthogonal direction has been chosen since the tracer particle displays a strong anomalous behavior in direction of the force, including intermittent superdiffusivity as well as a transition to a strongly enhanced long-time diffusion. This behavior has been intensively discussed in \cite{Schroer2013a}, where a quantitative theoretical description of the anomalous diffusion in terms of the PEL approach is provided.
The fact that the diffusion coefficient of the bath particles displays a nonlinear behavior has of course to be treated with some care, since it is only valid for the rather small system studied in this work, where the bath particles and the tracer belong to the same elementary system. Nevertheless, the analysis of this quantity may be instructive for the question of the universality of the effective temperature in the elementary system.

In analogy to the mobility in III.C., the nonlinear diffusion coefficients can be separated into its spatial and temporal compounds \cite{Doliwa2003c,Schroer2013}: 

\begin{equation}
  D_\text{X}(F,T) = \left(\av{a_\text{X}^2}/2\right)~\av{\Gamma(0,T_\eff)}_P,
 \label{eq:DF}
\end{equation}
where $\text{X}=\perp$ or $\text{X}=\text{A},\perp$ characterizes the different species. This equation is basically the same as \eq{eq:DT}, however, both the diffusive length scales $\av{a^2}$ and the average hopping rate $\av{\Gamma(T)}_P$ have been replaced by their force dependent counter parts. The only difference between for $D_\perp$ and $D_{A,\perp}$ is the relevant diffusive length scale, either $\av{a^2_\perp}$ or $\av{a^2_{\text{A},\perp}}$, which can be determined from the mean-squared displacements of the respective particle types after a certain number of MB transitions.

For the evaluation purely in real space, \eq{eq:DF} is less helpful since the PEL parameters are inaccessible. Similarly to what has been done for the mobility in the previous section, one one can therefore make the assumption that the diffusive length scales are basically independent of the external force (and temperature, since $\lim_{F\to0}\av{a^2_\text{X}(F,T)} = \text{const}$). If this assumption holds, \eq{eq:DF} simplifies to

\begin{equation}
 D_\text{X} \approx D(T_\eff)
 \label{eq:DF2}
\end{equation}
for both diffusion coefficients. The test of this assumption is shown in \fig{fig:FTSD}{(a),(b)}.

For the diffusion coefficient of the tracer particle one finds that the theoretical prediction is only in a qualitative agreement with the rescaled data points. Please notice, however, that in the analyzed temperature range the diffusion coefficient varies over three orders of magnitude, which is reasonably reflected by the theory. For $D_{\text{A},\perp}$, a 
much better agreement can be observed: The data points collapse on a master curve which agrees up to the harmonic breakdown at $T\approx 0.7$ with \eq{eq:DF2}. 

The reason for the different qualities of the prediction is the validity of the assumption that the spatial part is not dependent on the applied force. As one can see in \fig{fig:FTSD}{(c)}, the spatial part of the bath particles (lower branch) varies only very little as compared to the quiescent length scale. In contrast, the length scale of the tracer particle grows half an order of magnitude (upper branch). In the first case, the applied assumption is much better fulfilled, resulting in a better scaling in \fig{fig:FTSD}{(b)}.
For the tracer particle diffusion, please consider, that the spatial part of the tracer diffusion varies \emph{at most} half an order of magnitude. In the same regime, the temporal part, which can be very well described by our model, extends over one to two orders of magnitude (see \fig{fig:FTSmu}{(b)}). This explains why the presented approach still allows a reasonable description of the data in \fig{fig:FTSD}{(a)}.

Our results shown above are in some disagreement with the work presented in \cite{Winter2012,Winter2013} where it has been shown that the mobility and the orthogonal diffusion of the tracer particle obey a Force-Temperature Superposition, however, with different effective temperatures. Adopted to our notation, these two temperatures differ in the value of $\chi$ which turns out to be a factor of $\sim 1.3$ larger for the orthogonal diffusion than for the nonlinear mobility. Indeed, also for our data a larger value of $\chi$ (here increased by $70\%$ for a different system) allows a nearly perfect scaling of the numerical data on a master curve, see \fig{fig:FTSD}{(d)}. However, the effective temperature resulting from such a rescaling is somewhat artificial, since it has lost its connection to the thermodynamic and, in this application, kinetic properties it was originally derived from. Effective temperatures purely derived from scaling laws should therefore be treated with some caution, because its importance for the description of non-equilibrium thermodynamics is a priori unclear. 

\section{Conclusion}
In this work, we have derived a microscopic approach to describe the thinning behavior of nonlinear active microrheology in supercooled liquids. For this purpose we have applied the PEL approach to describe the thermodynamics and dynamics of the driven system which offers valuable insight into the relationship of each other, even far away from equilibrium. Our results could show that the thermodynamic and dynamic properties of the system far from equilibrium are governed by a single effective temperature that can qualitatively connected to the properties to the quiescent energy landscape. Since one can transfer this behavior, within some approximations however, to real space quantities like the force-dependent mobility and the orthogonal diffusion coefficient, it may encourage tests of this results in experiments as well as in other theoretical models of supercooled liquids.

\begin{acknowledgments}
This work was financially supported by DFG Research Unit 1394 "Nonlinear Response to Probe Vitrification" and NRW Graduate School of Chemistry. We furthermore acknowledge helpful discussions with E.~Bouchbinder, D.~Lacks, N.~H.~Siboni and L.~Smith.
\end{acknowledgments}

\end{document}